\documentclass[prb,aps,epsf,twocolumn,superscriptaddress,longbibliography]{revtex4-2}
\usepackage{graphicx}
\usepackage{dcolumn}
\usepackage{bm}
\usepackage{color}
\usepackage{soul}
\usepackage{hyperref}
\usepackage{amssymb}
\usepackage{amsmath}
\usepackage{wasysym}
\usepackage{tabularx}

\usepackage{appendix}
\usepackage{hyperref}
\usepackage{float}

\newcommand{\Uone}{\text{SU}(4)_1}
\newcommand{\SU}{\text{SU}(4)}

\begin{document}
\preprint{APS/123-QED}

\title{Charge-$4e$ Anyon Superconductor from Doping $\text{SU}(4)_1$ chiral spin liquid}

\author{Lu Zhang}
\affiliation{%
 Department of Physics, Hong Kong University of Science and Technology, Clear Water Bay, Hong Kong, China}

\author{Ya-Hui Zhang}
\affiliation{%
 Department of Physics and Astronomy, Johns Hopkins University, Baltimore, Maryland, USA 
}
\author{Xue-Yang Song}
\affiliation{Department of Physics, Hong Kong University of Science and Technology, Clear Water Bay, Hong Kong, China} 
\date{\today}

\begin{abstract}
Previous studies have shown that $\text{SU}(4)_1$ chiral spin liquid can emerge in the SU($4$) Hubbard model on triangular lattice. A natural question then arises: What is the phase upon doping? In this work, we show the possibility that hole doping can give rise to an anyon superconductor and propose that both spinons and holons form integer quantum Hall states with opposite chiralities.  Using topological field theory we demonstrate that the phase is a topological charge-$4e$ superconductor with chiral central charge $c_-=4$. We further identify the deconfined excitations and anyonic excitations bound to the vortex. This unusual superconductor may be realized in moir\'e bilayer and detected through quantized thermal Hall effect and quantum spin Hall effect.\\
\end{abstract}

\maketitle

\section{Introduction}
Since the discovery of the superconductor in 1911\cite{onnes1911superconductivity}, the exploration of the mechanisms behind the superconductivity is one of the long-standing topics in condensed matter physics. In addition to the BCS theory\cite{bardeen1957theory}, which describes the conventional superconductors through the condensation of the Cooper pairs, other theories have also been proposed to explain newly discovered superconductivity beyond the conventional mechanism.  
For example, the theory of anyon superconductor proposes that the superconductivity can emerge within an anyon gas\cite{laughlin1988superconducting,lee1989anyon,chen1989anyon}. 
However, it remains uncertain whether this phase can be stabilized in solid state systems. 
Recent experiment has observed the signature of the unconventional superconductivity in twisted $\text{MoTe}_2$ devices\cite{xu2025signatures} near the fractional quantum Hall phase.
 It is  suggested that the mechanism behind may be the anyon superconductivity\cite{shi2024doping}.  

We now turn our attention to the chiral spin liquid (CSL) state\cite{wen_zee_chiral},  an analog of the fractional quantum Hall phase in spin language. Numerous numerical and theoretical studies have proposed that an SU(2)$_1$ chiral spin liquid can yield an anyon superconductor\cite{tang2013superconductivity,ko2009doped,jiang2020topological,song2021doping,kim2025topological,divic2024anyon,shi2024doping}.
In this work, we investigate the possibilities of realizing an anyon superconductor phase by doping the $\Uone$ CSL in $\text{SU}(4)$ Hubbard model, where $\Uone$ refers to its field theory description and will be elaborated later. The low-energy excitation spectrum of CSL is gapped and the quasi-particles are anyons with self-statistics dictated by the underlying topological order. In particular, the $\Uone$ CSL considered here hosts anyons with self-statistics of $\pi/4$. 
The previous work\cite{zhang20214} has demonstrated that the $\Uone$ CSL can be stabilized in this model within a wide parameter regime at filling fraction $\nu = 1$(one electron at each site).
It is natural to ask what is the fate of the CSL upon doping.
Quantum Monte Carlo methods suffer from the sign problem and DMRG remains inaccessible to the doped $\text{SU}(4)$ Hubbard model regime; thus, there is so far no controlled, unbiased, numerical method to obtain the exact ground state.\\

Our study begins with the CSL phase at $\nu = 1$ and conducts a field theory analysis to explore the possible low-energy phases upon doping the CSL.
We adopt the slave particle (i.e., parton) approach and express the electron as $c^\dagger_{i\alpha}=f_{i\alpha}^\dagger b_i$, where $f^\dagger_{i\alpha}$ creates the spinon and $b_i$ annihilates the holon; $\alpha$ refers to the spin indices. It is known that the low-energy physics of the Hubbard model can be effectively described by the \textit{holon} and \textit{spinon} degrees of freedom when the interaction $U$ is strong. 
The ground state properties can be inferred through the interplay between the electrically charged holons and spin carrying spinons. 
For example, the $\Uone$ CSL can be understood as the $C=4$ Chern insulator of spinons while the holons are gapped out at $\nu = 1$, i.e. charge dynamics quenched.
Away from the integer filling, there are itinerant holes in the ground state.
To properly describe the low-energy physics, the holons degrees of freedom must be included.
One of the conventional stories of the doped Hubbard model is the \textit{holon condensation}, which proposes that holons condense while the spinons are in the spin liquid phase with the spinon Fermi surface. The $\text{U}(1)$ symmetry breaking of holons effectively results in a Fermi liquid state of the electrons through the Higgs mechanism \cite{song2021doping}.
However, it is still possible to realize a translation invariant state. In this phase, spinons still occupy a Chern band with the Chern number $C=4$, while the holons form the $C=-4$ bosonic quantum Hall phase(BQH) due to an internal flux shared between the spinons and holons.  
After integrating out the gapped matter fields(spinon and holon), the system is effectively described by the Chern-Simons field theory of the emergent gauge field. Based on $K$-matrix analysis, we demonstrate that the final phase is a charge-$4e$ superconductor with chiral central charge $c_-=4$.
Unlike in the case of doping an SU(2)$_1$ CSL, the superconductor described here is not connected to a conventional BCS state. The mechanism for charge-$4e$ pairing proposed here also differs from previous proposals that rely on vestigial order at finite temperature\cite{zhou2022chern,berg2009charge,jian2021charge,liu2023charge}.

\begin{table}
    \centering
    \renewcommand\arraystretch{1.2}
    \begin{tabular}{m{5em} m{1em} m{7em} m{1.5em} m{10em}}
    \hline\hline
    spinon    & + &  holon        & $\Rightarrow$ & electron  \\  \hline
     C=4 QH    & + &  C=-4 QH      & $\Rightarrow$ &  anyon superconductor  \\ 
    ~~SFS    & + &  Mott insulator           & $\Rightarrow$ &  ~~~~~CSL  \\
    ~~SFS    & + &  superfluid   & $\Rightarrow$ &  Fermi liquid  \\ 
    \hline\hline
    
    \end{tabular}
    \caption{The correspondence between the phases of parton(spinon and holon) and electrons at finite doping $x$. "QH" stands for the quantum Hall state and "SFS" for spinon fermi surface. The anyon superconductor can be viewed as the coupled “quantum Hall bilayer” with opposite Chern number in each layer. }
    \label{tab:parton_electron}
\end{table}

\section{Parton description of doping $\Uone$ chiral spin liquid}

We first introduce the SU($4$) Hubbard model on the triangular lattice, which is defined as
\begin{equation}
H = t\sum_{\langle  i 
 j \rangle} \left(c^\dagger_{i\alpha}c_{j\alpha} + \text{H.c.} \right)
 +\frac{U}{2}\sum_{i}n_{i}(n_{i}-1),
 \label{eq:Hubbard}
\end{equation}
where the repeated Greek index $\alpha=1\cdots 4$ is summed over. This could emerge as the effective model for double Moire bilayer systems\cite{zhang20214} or simulated on cold atom platforms\cite{yang2024chiral,heinz2020state,aidelsburger2013realization,cooper2019topological,gorshkov2010two,honerkamp2004ultracold,rapp2008trionic,wu2003exact}. At integer filling $\nu = 1$, the Hubbard model can be reduced to the $J-K$ model in the large-$U$ regime via the second order perturbation theory\cite{zhang2025phases}:
\begin{equation}
    H=J \sum_{\langle i j\rangle} S_i^{\alpha \beta} S_j^{\beta \alpha}+3K \sum_{i j k \in \Delta} S_i^{\alpha \beta} S_j^{\beta \gamma} S_k^{\gamma \alpha},
    \label{eq:JK}
\end{equation}
where $S^{\alpha \beta}$ represents the $\text{SU}(4)$ generators and the parameters $J = 2(t^2/U) -12(t^3/U^2)$ and $K = 2(t^3/U^2)$.
The ground state of the model is $\Uone$ CSL in parameter regime $U/t \in [12.2,24.4]$.
To theoretically describe the CSL, we adopt the parton formalism. Expressing in terms of parton, the $\text{SU}(4)$ generators $S^{\alpha \beta} = f^\dagger_\alpha f_\beta$.
In the CSL phase, the spinons form the $C=4$ Chern insulator.

To describe the system away from $\nu = 1$ we have to consider the itinerant holes by including the hopping term. Upon doping, the $J-K$ model is generalized to the $t-J-K$ model:
\begin{equation}
        H=t\sum_{\langle ij\rangle}Pc_{i\alpha}^\dagger c_{j\beta}P + J \sum_{\langle i j\rangle} S_i^{\alpha \beta} S_j^{\beta \alpha}+3K \sum_{i j k \in \Delta} S_i^{\alpha \beta} S_j^{\beta \gamma} S_k^{\gamma \alpha}.
    \label{eq:tJK}
\end{equation}
The operator $P$ is the Gutzwiller projector that rules out the double occupancy. From the low-energy perspective, the electrons are fractionalized into holons and spinons. The electron creation operator can be expressed as $c^\dagger_{i\alpha}=f_{i\alpha}^\dagger b_i$ with $f^\dagger_{i\alpha}$ creating the spinon and $b_i$ annihilating the holon. The parton decomposition introduces a gauge redundancy, as the physical electron operator is invariant under a phase rotation $b_i\rightarrow b_ie^{i\phi},f_{i\alpha}^\dagger\rightarrow f_{i\alpha}^\dagger e^{-i\phi}$. Hence a $U(1)$ gauge field $\alpha$ is introduced when going beyond mean-field treatment to include quantum fluctuations.
The gauge constraint from the parton decomposition reads:
\begin{equation}
    \hat n_i^b+\hat n_i^f = 1,
    \label{eq:constraint}
\end{equation}
where $\hat n_i^b = b^\dagger_i b_i$ and $\hat n_f = \sum_{\alpha}f^\dagger_\alpha f_\alpha$ are the holon density and spinon density operator. It is challenging to enforce this constraint exactly. Thus this equation is often treated on average: 
\begin{equation}
     \langle \hat n_i^b\rangle  +  \langle \hat n_i^f \rangle =1 . 
\end{equation}
To make further progress, we adopt the path integral to describe the low-energy physics. The partition function of the system is $\mathcal Z = \int \mathcal D[b,f,\chi,\mu] e^{-\mathcal S}$, where the action $\mathcal S$ takes the form:
\begin{widetext}
\begin{equation}
\begin{aligned}
    &\mathcal S= \int_\tau  \sum_{i}\left(|(\partial_\tau -\mu_i)b_i|^2+  
    f^\dagger_{i\alpha} (\partial_\tau-\mu_i) f_{i\alpha}\right)-   \sum_{\langle ij \rangle} \left(b_i^\dagger b_j f_{j\alpha}^\dagger f_{i\alpha}+f^\dagger_{j\alpha} f_{i\alpha}\chi_{ij} +\chi_{ij}\chi_{ji} \right)+\sum_{\langle ijk \rangle }  \chi_{ij}\chi_{jk}\chi_{ki},
\end{aligned}
\label{eq:partition_function}
\end{equation}
\end{widetext}
where $\mu_i$ is the Lagrangian multiplier to enforce the constraint\eqref{eq:constraint} and the $\chi_{ij} = \langle f^\dagger_i f_j \rangle$ is identified with the spinon hopping parameter.  
At integer filling $\nu=1$, the holons are gapped out.  
The CSL is described, at mean-field level for the spinons, by the hopping amplitudes on bond $\langle ij\rangle$ as $\chi_{ij}$ with $\pi/4$ flux per triangle, where spinons occupy the lowest four-fold degenerate Chern bands, with Chern number $C=4$.
The gauge field $\alpha$ couples to  the phase fluctuation of the order parameter, such that $\chi_{ij} = |\chi_{ij}|e^{i\alpha_{ij}}$. 

\begin{figure}
    \centering
    \includegraphics[width=1.0\linewidth]{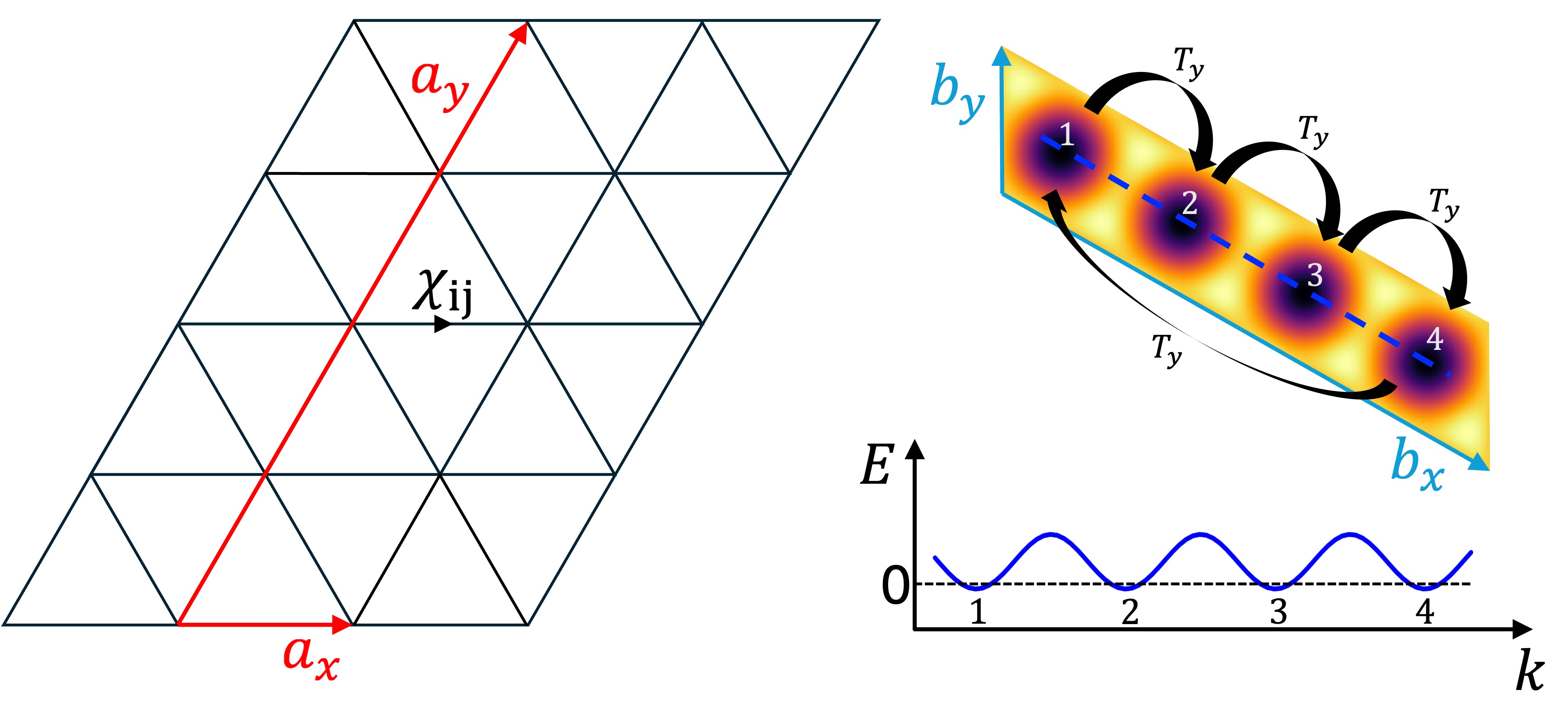}
    \caption{Left: The triangular lattice with magnetic unit cell defined by $a_x$ and $a_x$.  Top right: The magnetic Brillouin zone with four band minimum, named as $1$, $2$, $3$, $4$. Different band minimums are connected by the magnetic translation symmetry $T_y$. Bottom right: The holon band structure along the line cut shown in the magnetic Brillouin zone.}
    \label{fig:ansatz}
\end{figure}
With finite doping, the holon becomes indispensable to describe the low-energy physics. In this case, 
when the holon density $\langle \hat n_i^b\rangle  = x$ does not vanish, the density of spinons is fixed as $1-x$. 
This leads to three possibilities:
the spinons forming spin liquid with the spinon Fermi surface with holon condensed(I), holons gapped due to the disorder(II) and
spinons still fully occupying the spinon bands with holon forming the bosonic integer quantum Hall state(III). 
The three possibilities are summarized in table \ref{tab:parton_electron}.

In this study we focus on the last possibility(III).
For spinons to stay in the gapped quantum Hall phase at doping level $x$, the flux seen by the spinons should be reduced by $\delta b =  2\pi\frac{x}{4}$ according to the Streda formula $\delta n_i^f=C\delta b$($\delta b$ is the change of the flux seen by spinons,not to be confused with holon annihilation operator $b_i$). On the other hand the holon also feels the change of the flux $\delta b$.This is because the flux of the electron is equal to the subtraction between the holon and spinon flux due to the parton construction $c^\dagger = f^\dagger b $. The zero flux experienced by the electrons indicates that the flux of holon always equal to the flux of spinon.
In the next section we show the construction of the integer quantum Hall state of the holons and that it is consistent with the translation symmetry of the lattice.

\section{Bosonic quantum Hall of Holons}
In this section, we argue why the holons in the $\SU$ Hubbard model could be in the bosonic quantum Hall phase.
Suppose the flux piercing each triangles of $\chi_{ij}$ remains unchanged, which is $\pi/4$ as the undoped case and the flux experienced by holon is also $\pi/4$ going around the triangle. The translation symmetry is realized projectively in the representation of holons.

Consider a general case where the flux per unit cell is $2\pi\frac{p}q$. The translation symmetry is centrally extended to the magnetic translation symmetry with the commutation relation:
\begin{equation}
    T_xT_y = e^{i \frac{p}{q} 2\pi}T_yT_x.
    \label{eq:commutation}
\end{equation}
It is convenient to define the magnetic Brillouin zone and the magnetic unit cell with the translation symmetry operator defined as: $\tilde T_x = T_x $, $\tilde T_y = T_y^q $. This corresponds to enlarge the original unit cell $q$ times along $y$ direction in the Landau gauge. 
Because $[\tilde T_x,\tilde T_y]=0$ the eigenvector of magnetic translation symmetry can be labeled as $|\tilde k_x, \tilde k_y\rangle$ with $\tilde k_x$ and $\tilde k_y$ inside the magnetic Brillouin zone.
In the following we prove that the single-particle dispersion has to be $q$-fold degenerate inside the magnetic Brillouin zone to realize the projective representation of the translation symmetry.
We choose a state in the magnetic Brillouin zone diagonalizes the $T_x$ operator such that $T_x|\tilde k_x,\tilde k_y \rangle= e^{i\tilde k_x} |\tilde k_x,\tilde k_y \rangle$. 
To satisfy the relation \eqref{eq:commutation}, we require that 
$T_y|\tilde k_x,\tilde k_y \rangle= e^{i\tilde k_y} |\tilde k_x + \frac{p}{q}2\pi,\tilde k_y \rangle$. Thus, for an arbitrary state $|\tilde k_x, \tilde k_y\rangle$ the symmetry operator $T_y$ guarantees that the states $|\tilde k_x, \tilde k_y\rangle$, $|\tilde k_x+\frac{p}{q} 2\pi, \tilde k_y\rangle$, $|\tilde k_x+\frac{2p}{q}2\pi, \tilde k_y\rangle$ $\cdots$ $|p\ 2\pi, \tilde k_y\rangle$ share the same energy. 
Considering one energy minimum in the magnetic Brillouin zone, there are also three other $q-1$ energy minimum in the magnetic Brillouin zone connected by the translation operator $T_y$. 

In our case, the flux $\frac{2\pi}{4}$ of each unit cell experienced by the holon leads to the four energy minima in the holon bands. 
Thus the holons created by $b^\dagger$ occupies the four degenerate band minimum corresponding to $(0,0),(\frac{\pi}{4},0),(\frac{2\pi}{4},0),(\frac{3\pi}{4},0)$ in the Brillouin zone, labeled by $(k_x^m,0)$. We label the holon fields in band minimums as $\psi^{m}$ with $m = 1,2,3,4$ denoting four band minimums. 
Therefore, the holon sectors of the system can be described by the boson field $\psi^m$ with the four flavors in low-energy.

Under the magnetic translation symmetry different band minima will be permuted as shown in Fig.~\ref{fig:ansatz}. Because the fields $\psi^m(x)$ are expanded around the four band minima in the Brillouin zone they carry the non-trivial quantum number under $T_x$. Under $T_x$, $\psi^{m\dagger} \rightarrow e^{ik_x^m} \psi^{m\dagger};~~\psi^{m} \rightarrow e^{-ik_x^m} \psi^{m}$ while under $T_y$:  $\psi^{m\dagger} \rightarrow \psi^{m+1\dagger};~~\psi^{m} \rightarrow  \psi^{m+1}$.

Upon doping we assume that the flux seen by holon of each flavor is reduced by $2\pi\frac{x}{4}$. 
The system can be effectively described as the four flavor bosonic system with $x$ bosons in the presence of $-\frac{x}{4}$ flux of magnetic field in the continuum limit. Since holons carry positive charge there are repulsive interaction between holons resulting in the BQH. It is believed that the BQH phase is more energetically favorable than the condensation phase with repulsive interaction. This has been well-established in various theoretical and numerical study of bosonic system\cite{senthil2013integer,levin2008lattice,sterdyniak2015bosonic,he2015bosonic,furukawa2013integer,grass2014quantum}. A simple picture to describe the BQH is through the mutual flux attachment: one of the species of bosons is attached with the flux seen by other species of bosons\cite{senthil2013integer}. To facilitate the flux attachment, we introduce the gauge field $\tilde \beta_m$ with Chern-Simons term that is minimally coupled with the boson field $\psi^m(x)$ such that
\begin{equation}
\mathcal L_{\text{BQH}} =\sum_m\psi^{m\dagger}(x)(\partial+\tilde \beta_m+A)\psi^m(x)  -\frac{1}{4\pi}\tilde \beta^m K^b_{mn} d\tilde \beta^n  ,
\end{equation}
where $K^{b}_{mn}$ is the $K$-matrix of the Chern-Simons theory.
The condensation of the composite boson locks the gauge field $\tilde \beta_m$ and $A$. We introduce the Lagrangian multiplier $\beta_m$
\begin{equation}
\mathcal L_{\text{BQH}} =  -\frac{1}{4\pi}\tilde \beta^m K^b_{mn} d\tilde \beta^n +\frac{1}{2\pi} \beta_m\left(-d\tilde \beta^m+q^m_cdA\right),
\end{equation}
where $\boldsymbol q_c =(1,1,1,1)$ is the charge vector. After integrating out the dual field $\beta_m$, $\tilde \beta^m$ is coupled with $q_c^mA$. To describe an bosonic system the diagonal term of the $K-$matrix should be even number. According to the transformation of field $\tilde \psi^{\dagger m}(x)$ under the magnetic translation symmetry, the gauge fields $\tilde \beta_m$ coupled with the field $\tilde\psi^m$ will be permuted correspondingly as $\tilde \beta_m \rightarrow \tilde \beta_{m+1}$ while unchanged under $T_x$. 
Integrating out the gauge field $\tilde \beta_m$($\beta^m$ is identified as $\beta_m$ in the following) leads to the Lagrangian
\begin{equation}
\begin{aligned}
    \mathcal L_{\text{BQH}} = & -\frac{1}{4\pi}\beta^m K^b_{mn} d\beta^n+\frac{1}{2\pi}\left(q_c^m\beta_m\right)dA+\cdots.
\end{aligned}
\end{equation}
We also require that our field theory satisfying the magnetic symmetry of microscopic Hamiltonian. 
To construct the magnetic translationally invariant states, we adopt the most straightforward Chern-Simons formulation expressed as $\mathcal L_{\text{BQH}}=\beta_1d\beta_3+\beta_2d\beta_4$. Under the permutation of the gauge field $\beta_m$, this Lagrangian density is invariant under the magnetic translation symmetry $T_x$, $T_y$. The $K$-matrix of the holon takes the following form: 
\begin{equation}
   K^b = \left(\begin{array}{cccc}
0 & 0 & -1 & 0 \\
0 & 0 & 0 & -1 \\
-1 & 0 & 0 & 0 \\
0 & -1 & 0 & 0
\end{array}\right),
\label{eq:K_BQH}
\end{equation}
which reveals that there are mutual flux attachment of first(second) and third(fourth) species of holons.  It can be verified that the Hall conductivity $\sigma_H=\boldsymbol{q}_c^T(K^b)^{-1}\boldsymbol{q}_c = -4$.  

There is a simple argument to show the emergence of the superconductivity phase.
According to the Ioffe-Larkin rule, the resistivity of the system $\rho$ can be expressed as the summation of  resistivity of spinons $\rho^f$ and holon $\rho^h$: $\rho = \rho^f+\rho^h$\cite{ioffe1989gapless,chen2024non}. According to previous discussion, the holons and spinons are all in the quantum Hall phases with opposite Chern number such that $\rho^h = \left( \begin{array}{cc} 0 & -4 \\
4 & 0\end{array} \right) $ and $\rho^f = \left( \begin{array}{cc} 0 & 4 \\
-4 & 0\end{array} \right) $. Thus the resistivity of the system $\rho$ vanishes, which indicates that the system is in the superconductor phase. 
It should be noted that the spontaneous breaking of time-reversal symmetry is crucial for the formation of anyon superconductor. Consider the Hofstadter Hubbard model that breaks the time-reversal symmetry explicitly by threading flux in the unit cell. The holon would see the external magnetic flux in addition to the emergent flux. Thus it is not guaranteed that there is a four-fold degeneracy of the holon band, which leads to the breakdown of the arguments.
To enhance the above arguments, we go beyond the mean-field approach and obtain the topological field theory of the system\cite{fisher1989correspondence,blok1990effective}.

\section{Emergence of anyon superconductor}
\label{sec:anyonsc}

The Lagrangian of the  CSL in terms of gauge field $\alpha_i$'s can be obtained by integrating out the spinon matter field $f_i^\dagger$ and the holon matter field:
\begin{equation}
\begin{aligned}
\label{eq:l_f}
    \mathcal L_{\text f} = -\frac{1}{4\pi}\sum_{i=1}^4\alpha_i d\alpha_i-\frac{1}{2\pi}ad\left(\sum_{i=1}^4 \alpha_i\right)+\cdots  ,
\end{aligned}
\end{equation}
where $ada$ is the shorthand notation for $\epsilon_{\mu\nu\rho}a^\mu \partial^\nu a^\rho$ and and the $4$ auxiliary fields $\alpha_i$ is introduced to descirbe the Chern insulator of the spinons with $C=4$ .
The effective field theory for holon degrees of freedom reads
\begin{equation}
\begin{aligned}
    \mathcal L_{\text{b}} = & \mathcal L_{\text{BQH}} + \frac{1}{2\pi}ad\left(\sum_{m=1}^4\beta_m\right)+\cdots
\end{aligned}
\end{equation}
after we integrate the gapped matter field $b$.   The gauge field $a$ is the $U(1)$ gauge field that comes from the parton decomposition and couples to holons and spinons. $\cdots$ represents the Maxwell term, which is irrelevant in the presence of the Chern-Simons term.

\begin{table*}
    \centering
    \renewcommand\arraystretch{1.5}
    \begin{tabular}{m{8em} m{12em} m{14em}m{4em}m{4em}}
    \hline \hline
    anyon types    &$(q_1, q_2, q_3, q_4, q_5, q_6, q_7)$ & $\ (\tilde q_1, \tilde q_2, \tilde q_3, \tilde q_4, \tilde q_5, \tilde q_6, \tilde q_7)$ &  $\theta$ &  $V$      \\ [0.2ex] \hline
     $\quad$ type I       & $\ (1,\ 0,\ 0,\ 0,\ 0,\ 0,\ 0)$ & $(1,-1,-1,-1,\ 0,\ 0,\ 0)$ & $\frac{3\pi}{4}$ & 1    \\
    $\quad$ type II        & $\ (0,\ 1,\ 0,\ 0,\ 0,\ 0,\ 0)$ &  $(1,\ \ 1,-1,\ \ 1,\ 0,\ 0,\ \ 0)$  & $\frac{3\pi}{4}$ & 1     \\
    $\quad$ type III        & $\ (0,\ 0,\ 1,\ 0,\ 0,\ 0,\ 0)$ & $(1,-1,\ \ 1,\ \ 1,\ 0,\ 0,\ \ 0)$ & $\frac{3\pi}{4}$ & 1      \\
    $\quad$ type IV        & $\ (0,\ 0,\ 0,\ 1,\ 0,\ 0,\ 0)$ & $(1,\ \ 1,\ \ 1, -1,\ 0,\ 0,\ \ 0)$ & $\frac{3\pi}{4}$ & 1      \\
    $\quad$ type V        & $\ (0,\ 0,\ 0,\ 0,\ 1,\ 0,\ 0)$ & $(1,\ \ 0,\ \ 0,\ \ 0,\ 1,\ 1,\ \ 1)$ & $\frac{\pi}{4}$ & 1      \\
    $\quad$ type VI       & $\ (0,\ 0,\ 0,\ 0,\ 0,\ 1,\ 0)$ & $(1,\ \ 0,\ \ 0,\ \ 0,\ 1,\ 0,-1)$ & $\frac{\pi}{4}$ & 1           \\
    $\quad$ type VII        & $\ (0,\ 0,\ 0,\ 0,\ 0,\ 0,\ 1)$ & $(1,\ \ 0,\ \ 0,\ \ 0,\ 1,-1,\ \ 0)$ & $\frac{\pi}{4}$ & 1      \\
         \hline \hline
    \end{tabular}
    \caption{List of quasi-particles and their mutual statistics. $\theta$ is the self-statistics and $V$ is the vortex charge, which is the $\tilde q_1$ component. We find that the quasi-particles with fractional statistics in the list carry the vortex charge. $q_i$ labels the charge in the original basis and $\tilde q_i$ labels the charge in the basis after the transformation.  }
    \label{tab:quasi-particles}
\end{table*}

To simplify the expression we first integrate out the gauge field $a$ leading to the constraint:
\begin{equation}
    \sum_i^4 \alpha_i= \sum_m^4\beta_m.
\end{equation} 
Then we can eliminate the gauge field $\beta_4$ by expressing $\beta_4 = \sum_i \alpha_i - \sum_{1\leq m \leq 3} \beta_m$. For convenience, we define the 
new notation such that $\alpha_i = \alpha_i(1\leq i \leq 4)$ and $\alpha_i = \beta_{i-4}(4< i \leq 7)$. The first four flavors represent the four spinon degrees of freedom and the last three represent the independent holon degrees. 
In a more compact form, the Lagrangian can be written as
\begin{equation}
    \mathcal L = -\frac{1}{4\pi}  \alpha_i K^{ij}  d \alpha_j+
    \frac{1}{2\pi}A_cd(\sum_i^7 \alpha_i),
\end{equation}
where we use the notation such that $\alpha_i = \alpha_i(1\leq i \leq 4)$ and $\alpha_i = \beta_{i-4}(4< i \leq 7)$ as in the main body. The first four components represents the holon excitations and the last three components represent the spinon excitations. The $K$-matrix in the basis can be expressed as:
\begin{equation}
   K = \left(\begin{array}{ccccccc}
1 & 0 & 0 & 0 & 0 & -1 & 0 \\
0 & 1 & 0 & 0 & 0 & -1 & 0 \\
0 & 0 & 1 & 0 & 0 & -1 & 0 \\
0 & 0 & 0 & 1 & 0 & -1 & 0 \\
0 & 0 & 0 & 0 & 0 & 1 & -1 \\
-1 & -1 & -1 & -1 & 1 & 2 & 1 \\
0 & 0 & 0 & 0 & -1 & 1 & 0
\end{array}\right).
\end{equation}
This $K$-matrix contains an eigenvector $\boldsymbol{q}_0 = (1,1,1,1,1,1,1)^T$  with zero eigenvalue. This null eigenvector implies that the Chern-Simons terms corresponding to this vector vanishes. The lowest-order term is the Maxwell term and the the excitation is gapless. It is thus identified with the superfluid mode.  The charge vector is $\boldsymbol{q}_c=(1,1,1,1,0,0,0)^T$. The non-zero inner product $\boldsymbol{q}_0 \cdot \boldsymbol{q}_c$ indicates that the superfluid mode is coupled with the external electromagnetic field.
Because the null vector $\boldsymbol{q}_0$ carries four holons, each of which carries an unit charge the superfluid mode carries four unit charges. This leads to one of the conclusions that the superconductor resulting from doping the $\Uone$ chiral spin liquid is charge-$4e$.
Furthermore, the superconductor from doping the $\Uone$ CSL support the chiral edge mode with chiral central charge $c_- = 4$. This is derived from the difference between the numbers of positive and negative eigenvalues of the $K$-matrix\cite{lu2016classification}.

We relabel the gauge field $\tilde \alpha_i = T^{ij}\alpha_j$ to express the action into a more transparent form. 
We choose the new basis such that (I) the first basis vector is the null vector, representing the goldstone mode of superfluid. It is required that the minimal vortex in superconductor phase is $\frac{2\pi}{4}$. (II) The remaining $6$ basis vectors are orthogonal to the null vector. Physically, this means that these modes are decoupled from the superfluid mode.
Thus, we define the transformation matrix $T$ as
\begin{equation}
    \begin{aligned}
        T=\left(\begin{array}{ccccccc}
1/4 & 1/4 & 1/4 & 1/4 & 0 & 0 & 0 \\
-1/4 & 1/4 & -1/4 & 1/4 & 0 & 0 & 0 \\
-1/4 & -1/4 & 1/4 & 1/4 & 0 & 0 & 0 \\
-1/4 & 1/4 & 1/4 & -1/4 & 0 & 0 & 0 \\
-1/4 & -1/4 & -1/4 & -1/4 & 1/3 & 1/3 & 1/3 \\
0 & 0 & 0 & 0 & 1/3 & 1/3 & -2/3 \\
0 & 0 & 0 & 0 & 1/3 & -2/3 & 1/3
\end{array}\right).
    \end{aligned}
\end{equation}
The action written in the field $\tilde \alpha_i$ is:
\begin{equation}
    \mathcal{L}=-\frac{1}{4 \pi} \tilde \alpha_i \tilde{K}^{i j} d \tilde \alpha_j+\frac{4}{2\pi}A_cd\tilde \alpha_1,
    \label{eq:Action in tilde alpha}
\end{equation}
where the $K$-matrix in the new basis is:
\begin{equation}
    \tilde K = \left(\begin{array}{ccccccc}
0 & 0 & 0 & 0 & 0 & 0 & 0 \\
0 & 4 & 0 & 0 & 0 & 0 & 0 \\
0 & 0 & 4 & 0 & 0 & 0 & 0 \\
0 & 0 & 0 & 4 & 0 & 0 & 0 \\
0 & 0 & 0 & 0 & 4 & -4 & 0 \\
0 & 0 & 0 & 0 & -4 & 2 & 1 \\
0 & 0 & 0 & 0 & 0 & 1 & 0
\end{array}\right)
\end{equation}
The zero element of the above $K$-matrix signals the absence of the Chern-Simons term of the $\tilde \alpha_1$. The only term involving $\tilde \alpha_1$ in the Lagrangian is $\frac{4}{2\pi}A_cd\tilde \alpha_1$, which can be viewed as the Goldstone mode and will higgs the gauge field $A_c$, leading to the Meissner effect of the resulting superconductor.

We emphasize that this construction in fact realizes the celebrated anyon superconductivity proposed decades ago \cite{lee1989anyon}. Upon doping the $\Uone$ chiral spin liquid, a dilute gas of anyons with self-statistics of $\pi/4$ and unit electric charge is obtained. The mechanism of anyon superconductivity hence leads to a superfluid of charge-$4e$ objects, i.e. charge-$4e$ superconductivity.  From the viewpoint of the holons, the part of Lagrangian resulting from integrating out spinons $\mathcal L_f$ (eq \eqref{eq:l_f}) realizes the flux attachment:integrating out $\alpha_i$'s in eq \eqref{eq:l_f}, we have
\begin{align}
    \mathcal L=\mathcal L_{b;a}+\frac{4}{4\pi} ada,
\end{align}
where the first part describes the field theory for holons $b$ that couples to the gauge field $a$. The equation of motion from taking derivative on $a_0$ reads $\rho_b=4 \nabla\times \mathbf a/(2\pi)$, which implements flux attachment and transmutes each holon $b$ to an anyon with self-statistics of $\pi/4$. This furnishes the description of an anyon gas with self-statistics of $\pi/4$. Our analysis in section \ref{sec:anyonsc} therefore links the long-wavelength theory upon doping $\Uone$ chiral spin liquid to the effective theory of anyon superconductivity proposed in Ref\cite{lee1989anyon}. 
The $4e$ Cooper pair could be visualized most directly from the following intuitive picture: one unit of flux quantum of $a$ nuleates $4$ spinons and $4$ holons from their Hall response, respectively, forming the superconducting pairs.

After the transformation from the gauge field $\alpha$ to $\tilde \alpha$ the charge quantization should be changed correspondingly,
\begin{equation}
    \tilde{\boldsymbol{q}}_i = (T^{T})^{-1} \boldsymbol{q}_i
\end{equation}
with the null vector $\tilde{\boldsymbol{q}}_1$ in the new basis. Thus the excitation can be labeled by a vector $\tilde{\boldsymbol{q}}=[\tilde q_1,\tilde q_2, \cdots, \tilde q_7]$. The subtlety is that not all the $\tilde{\boldsymbol{q}}$ is the true excitation in the system. To satisfy the charge quantization, the quasi-particles in the new basis must be composed of integer composition of the quasi-particles in the original basis.  The correspondence between the particles in two sets of basis can be read from the table~\ref{tab:quasi-particles}.
The self- and mutual-statistics of the quasi-particles can also be read out from the inverse of the $K_r$-matrix after we exclude the null vector:
\begin{equation}
        \tilde K^{-1}_r = \left(\begin{array}{cccccc}
1/4 & 0 & 0 & 0 & 0 & 0 \\
0 & 1/4 & 0 & 0 & 0 & 0 \\
0 & 0 & 1/4 & 0 & 0 & 0 \\
0 & 0 & 0 & 1/36 & 1/9 & -2/9 \\
0 & 0 & 0 & 1/9 & 4/9 & 1/9 \\
0 & 0 & 0 & -2/9 & 1/9 & -2/9
\end{array}\right)
\end{equation}

The information of the quasi-particles is encoded in the $K$-matrix and the charge vector. For example, for particles $\boldsymbol{q}_1$ and $\boldsymbol{q}_2$, its mutual statistics is defined as $2\pi\boldsymbol{q}_1 \tilde K^{-1}_r \boldsymbol{q}_2 $ and self-statistics is $\pi\boldsymbol{q} \tilde K^{-1}_r \boldsymbol{q}$. We list the quasi-particles excited in the holon spinon degrees of freedom in the table \ref{tab:quasi-particles} with self-statistics and vortex charge. In a superconductor, the vortices are confined. In the action \eqref{eq:Action in tilde alpha} the Chern-Simon term $\frac{4}{2\pi}A_c d\tilde{\alpha}_1$ indicates that $\tilde \alpha_1$ is the gauge field that coupled with the vortex charge. All the excited particles have the vortex charge,hence are confined. In addition to the seven types of particles listed in the table \ref{tab:quasi-particles}, there also exist  composite particles. For example type I and type V particles can combine into a new type of particle without vortex charge. 
These composite quasi-particles are actually "excitons" that carry different quantum numbers.
Since the vortex charge of these different quasi-particles is zero, these are deconfined excitation. For example, the holon and spinon combines to form the original electron degrees of freedom. From the $K$-matrix derived above, it is verified that it satisfies the fermionic degrees of freedom, which is consistent with the electron statistics. While the other composite quasi-particles are the bosons. 

\section{Discussion and conclusion}
In this study, we propose the emergence of an anyon superconductor by doping the $\Uone$ CSL. 
 We adopt the parton approach and topological field theory to show that the final phase is a charge-$4e$ superconductor. Our study provides another theory to realize the charge-$ne$ superconductor in realistic material in addition to previous study\cite{zhou2022chern} and can be detected experimentally through the thermal Hall coefficient $\kappa_{xy} = \frac{\pi T c_- }{6}$ with $c_-=4$. The spin Hall coefficient $\sigma_{xy}^s$ can also be obtained from the effective field theory\cite{zhang20214}. We hope our work can inspire future DMRG or other unbiased numerical tools to calculate the ground state of the doped Hubbard or $t-J-K$ model presented above. We also call for the future experimental study to explore the anyon superconductivity in the mori\'e system by doping the incompressible phase.\\

\section{Acknowledgments}

We thank Ashvin Vishwanath for previous collaboration and discussions. YHZ is supported by the National Science Foundation under Grant No. DMR2237031.
Lu Zhang is supported by Early Career Scheme of Hong Kong Research Grant Council with grant No. 26309524 and startup fund at HKUST

\bibliography{Main}

\onecolumngrid 

\appendix

\section{SU($N$) anyon superconductor}
The above derivation can also be extended to the SU($N$) case, where $N$ is an even number. $\text{U}(1)_N$ chiral spin liquid has been shown to be stabilized in $\text{SU}(N)$ Hubbard model in square and triangular lattice\cite{hermele2009mott,yao2021topological}. In the low-energy, the $\text{U}(1)_N$ CSL is described by the spinons degrees of freedom hopping on the lattice such that
\begin{equation}
    H = \sum_{\langle ij \rangle} \chi_{ji} f^\dagger_{i\alpha} f_{j\alpha}  + \text{H.c.} -\sum_i\mu_i f^\dagger_{i\alpha}f_{i\alpha},
\end{equation}
where $\chi_{ij} = \sum_{\alpha}\langle f^\dagger_{i\alpha}f_{j\alpha}\rangle/N$.
In the ground state, the spinons occupy the lowest Chern bands with Chern number $C=N$. 
The numerical result shows that the flux piercing each plaquette is $2\pi/N$. 
The story of doping the $\text{U}(1)_N$ CSL is the same as doping the $\Uone$ CSL.
Upon doping the holon experience $2\pi/N$ flux per unit cell. According to the projective translation symmetry arguments in the main text there are $N$ band minimum of boson, resulting in $N$ flavors. We assume that the holons form the bosonic quantum Hall phase as in the $\SU$ case in the low-energy by the change of flux experienced by holons. Then the spinons are still in the $C=N$ quantum Hall phase.
To describe the bosonic quantum Hall phase we have to introduce the emergent gauge field $\beta_m$, with $m = 1,2,3,\cdots,N$.
The effective theory of the SU($N$) case is in the same form as the $\SU$ case except the summation is from $1$ to $N$. While the $K$-matrix of the QBH phase is
\begin{equation}
    (K^b)^{ij}= \begin{cases}
    -1 &,   |i-j|=N/2 \\ 
    0   &,  \text{else}
    \end{cases}
\end{equation}
where $i$ takes from $1$ to $N$. The charge vector is $\boldsymbol{q}_c=(1,1,\cdots,1)$. When $N=4$, the $K$-matrix of the BQH reduces to the $K$-matrix \eqref{eq:K_BQH} in the main text.  The Hall conductivity can be calculated as $\sigma_H=\boldsymbol{q}_c^T(K^b)^{-1}\boldsymbol{q}_c = -N$.  Integrating the gauge field $a$ out leads to the constraint:
\begin{equation}
    \sum_i^N \alpha_i = \sum_m^N \beta_m
\end{equation}
Similarly, we can eliminate $\beta_N$ by inserting $\beta_N = \sum_i^N\alpha_i - \sum_m^{N-1}\beta_m$ into the Lagrangian. We obtain the following effective action:
\begin{equation}
    \mathcal{L}=-\frac{1}{4 \pi} \alpha_i K^{i j} d \alpha_j+\frac{1}{2 \pi} A_c d\left(\sum_i \alpha_i\right)
    \label{eq:SUNLagrangian}
\end{equation}
where $\alpha_i \equiv \alpha_i(1\leq i \leq N)$ and $\alpha_i = \beta_{i-N}(N<i\leq N-1)$.  The $K^{ij}$ of the Lagrangian \eqref{eq:SUNLagrangian}  takes the following form: 
\begin{equation}
\begin{aligned}
    K = \left( \begin{array}{ccc|cccccccc}
     1 &        &   &    &           &    &   -1     &    &         &  \\
       & \ddots &   &    &           &    &  \vdots  &    &         &  \\
       &        & 1 &    &           &    & -1       &    &         &  \\ \hline
       &        &   &    &           &    &   1      & -1 &         &  \\
       &        &   &    &           &    &   \vdots &   &  \ddots  &  \\
       &        &   &    &           &    &   1      &   &          & -1\\
    -1 &   \cdots   &  -1 &  1  &  \cdots   &  1  &   2      & 1  &     \cdots     & 1 \\
       &        &   & -1 &           &    &   1      &   &          &  \\
       &        &   &    &   \ddots  &    & \vdots   &   &          &  \\
       &        &   &    &           & -1 &  1       &   &          &  \\
    \end{array} \right)
\end{aligned}
\end{equation}

It can be verified that the determinant of the K  $\det(K)=0$, which means there is a zero mode contained in the $K$-matrix. In the following we are going to transform the $K$-matrix in a set of new basis to show the existence of the zero mode. There is an arbitrariness to transform the $K$-matrix to make the zero mode apparent. There is a simple solution that the transformation matrix $T$ can be expressed as:
\begin{equation}
    \begin{aligned}
        T=\left(\begin{array}{ccc|ccc}
 &  &    &  &  &  \\
 &  &    &  &  &  \\
 & \boldsymbol T^{NN} &    &  & O^{(N-1)N} &  \\
 &  &    &  &  &  \\
 &  &    &   &  &  \\ \hline
-1/N & \cdots & -1/N & 1/(N-1) & \cdots & 1/(N-1) \\ \hline
 &  &    &  &  &  \\
 &  &    &  &  &  \\
 & O^{N(N-1)} &    &   & \boldsymbol T^{(N-1)(N-2)}  & \\
 &  &    &  &  &  \\
 &  &    &  &  & 
\end{array}\right).
    \end{aligned}
\end{equation}
such that each element of $\boldsymbol{T}^{NN}$ should be $\pm 1/N$. The first row of $\boldsymbol{T}^{NN}$ is $\left(1/N ,\cdots,1/N\right)$. The remaining $(N-1)$ rows of $\boldsymbol{T}^{NN}$ should be the combination of $1/N$ and $-1/N$ such that the summation of the element of each row equal to $0$ and they are linearly independent. For $\boldsymbol{T}^{(N-1)(N-2)}$, each row contains one element equal to $-(N-2)/(N-1)$ and the remaining elements are $1/(N-1)$.
The inverse of the $T$ matrix always takes the form:
\begin{equation}
    T^{-1}=\left(\begin{array}{cccc}
1 & 1 & \cdots & 1 \\ \hline
  & &\cdots &\\
\vdots & & \ddots &    \vdots\\
 & & \cdots&
\end{array}\right)
\end{equation}
where the first row is the identity vector. It can be verified that the after the transformation the $K$-matrix takes the form:
\begin{equation}
    \begin{aligned}
       \tilde{K}  = \left( \begin{array}{c|ccc}
           0        &   0 &\cdots &  0\\ \hline
           0       &     &       &  \\
           \vdots   &     &       &  \\
           0        &     &       &
        \end{array}
        \right)
    \end{aligned}
\end{equation}
The $0$ block indicates the existence of the zero mode. After the transformation, The Lagrangian is:
\begin{equation}
    \mathcal{L}=-\frac{1}{4 \pi} \tilde{\alpha}_i \tilde{K}^{i j} d \tilde{\alpha}_j+\frac{4}{2 \pi} A_c d \tilde{\alpha}_1,
\end{equation}
where $\tilde \alpha_1$ coupled with the external gauge field is the Goldstone mode. It proves the existence of the superconductivity.
\end{document}